**Title:** Human sperm steer with second harmonics of the flagellar beat

**Authors:**

Guglielmo Saggiorato[1,#], Luis Alvarez[2,*,#], Jan F. Jikeli[2], U. Benjamin Kaupp[2], Gerhard Gompper[1], and Jens Elgeti[1,*]

[1]Institute of Complex Systems and Institute for Advanced Simulation, Forschungszentrum Jülich, 52425 Jülich, Germany. [2]Department of Molecular Sensory Systems, Center of Advanced European Studies and Research (CAESAR), 53175 Bonn, Germany.

[#]G.S. and L.A. contributed equally to this work.

[*]To whom correspondence should be addressed.
E-mail: j.elgeti@fz-juelich.de (J.E.); luis.alvarez@caesar.de (L.A.)

**Abstract:**

Sperm are propelled by bending waves travelling along the flagellum. During steering in gradients of sensory cues, sperm adjust the flagellar beat waveform. Symmetric and asymmetric beat waveforms produce straight and curved swimming paths, respectively. Two different mechanisms controlling the flagellar beat have been proposed: average intrinsic curvature and dynamic buckling instability. Both mechanisms create spatially asymmetric waveforms that could be modulated for steering. Using video microscopy, we image the flagellar waveform of human sperm tethered with the head to a glass surface. The waveform is characterized by a fundamental beat frequency and its second harmonic. We show that superposition of first and second harmonics breaks the beat symmetry temporally rather than spatially. As a result, sperm rotate around the tethering point. The rotation velocity is determined by the amplitude and phase of the second harmonic. Sperm stimulation with the female sex hormone progesterone enhances the second-harmonic contribution, modulates the flagellar beat, and ultimately sperm rotation. The temporal breaking of beat symmetry represents a new mechanism of sperm steering. Higher-frequency components were also reported for the flagellar beat of other cells; therefore, this steering mechanism might by quite general and could inspire the design of synthetic microswimmers.

Many microorganisms and cells are propelled by motile flagella or cilia, i.e. hair-like protrusions that extend from the cell surface[1-3]. The beat patterns of flagella vary among cells. The green algae *Chlamydomonas reinhardtii* navigates by a breaststroke-like movement of a pair of flagella that alternate between in-phase and anti-phase states of the beat[4]. The bacterium *Escherichia coli* is propelled by the rotary movement of a helical bundle of flagella; when the rotary direction reverts, the flagellar bundle disentangles and the cell randomly adopts a new swimming direction[5]. Most animal sperm swim by means of bending waves travelling from the head to the tip of the flagellum. Near a surface, sperm swim on a curvilinear path[6,7], which is thought to result from the spatial asymmetry of bending waves. Two different mechanisms have been proposed that could generate a flagellar asymmetry: a dynamic buckling instability resulting from flagellar compression by internal forces[8,9] or an average intrinsic curvature[6,10,11].

The flagellum not only propels sperm, but also serves as antenna that integrates sensory cues as diverse as chemoattractant molecules, fluid flow, or temperature that modify the flagellar beat pattern. Gradients of such chemical or physical cues guide sperm to the egg[12,13]. The sensory stimulation gives rise to changes in intracellular $Ca^{2+}$ concentration ($[Ca^{2+}]_i$) that modulates the flagellar beat and, thereby, swimming direction[13-17]. However, it is not known whether an intrinsic flagellar curvature, or a buckling instability, or some other mechanism is used for steering and how sensory cues modulate any of these mechanisms.

Here we study the flagellar beat pattern of tethered human sperm by high-speed, high-precision video microscopy and by theoretical and computational analysis. We show that the beat pattern is characterized by a superposition of two bending waves – with a fundamental frequency and its second harmonic – travelling down the flagellum. The second harmonic breaks the symmetry of the overall waveform by a temporal rather than by a spatial mechanism. The sexual hormone progesterone, which evokes a $Ca^{2+}$ influx, enhances the second-harmonic contribution and changes the rotation velocity of the cell. Our analysis suggests a novel temporal steering mechanism of sperm and uniflagellated eukaryotic microswimmers in general.

## Results

**Two harmonics shape the flagellar beat.** In a narrow recording chamber, we monitored the flagellar beat of human sperm (Fig. 1a). While swimming near a surface, sperm undergo a rolling motion. We prevent rolling by tethering sperm with their head to the recording chamber. Near a surface, the beat pattern is almost planar and parallel to the surface[8], which facilitates tracking and imaging of the flagellar motion. Sperm revolve around the tethered head with a rotation velocity $\Omega$ that varied smoothly over time between 0 and 0.5 Hz (Fig. 1a,b). The flagellar shape, extracted by image processing, was characterized by the local curvature $C(s,t)$ at time $t$ and arclength $s$ along the flagellum.

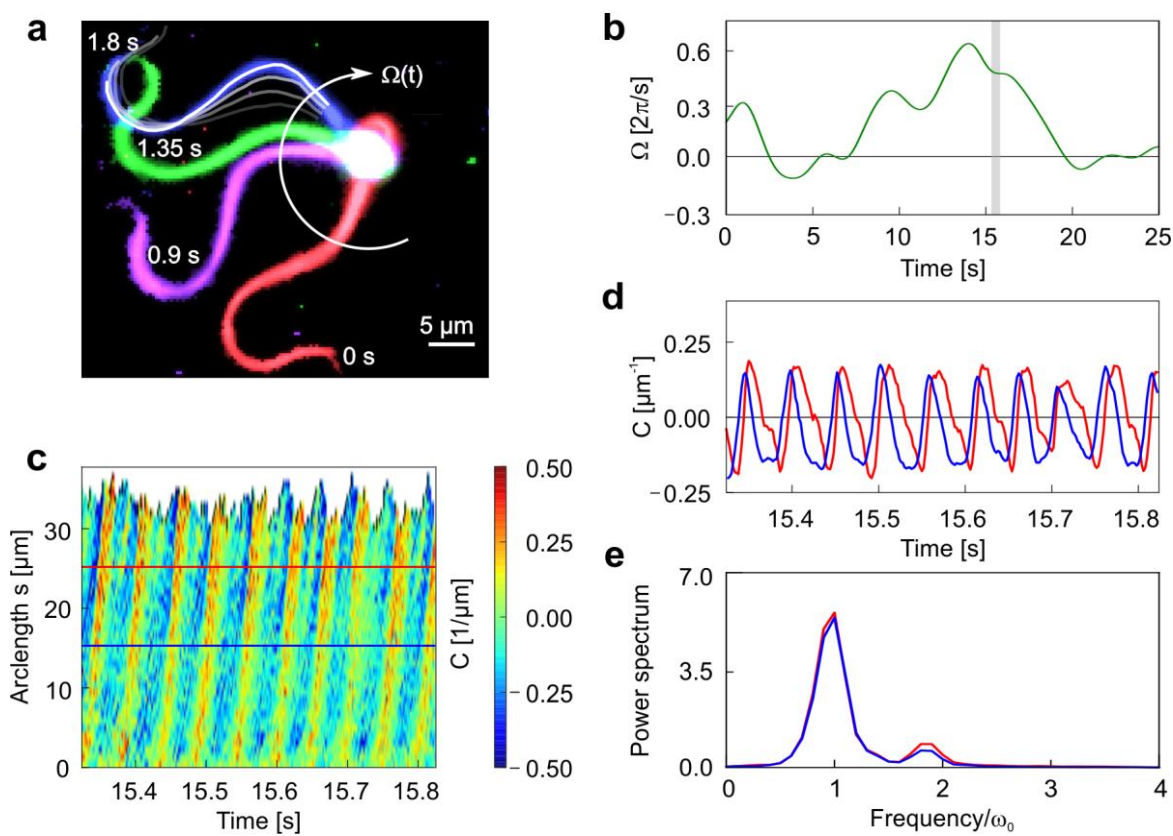

**Figure 1 | The flagellar beat pattern of human sperm displays a second-harmonic component.** (**a**) Four snapshots of a tethered human sperm that rotates clockwise around the tethering point with rotation velocity $\Omega(t)$. Each color corresponds to a different snapshot taken at the indicated time. White and grey lines below the blue snapshot show the tracked flagellum at consecutive frames acquired at 2 ms intervals. (**b**) Rotation velocity of the cell around the tethering point. The grey area indicates the time interval that is further analyzed in panels **c** and **d**. (**c**) Kymograph of the flagellar curvature during approximately 10 beat cycles (0.5 s). The curvature corresponding to the two horizontal lines is plotted in panel **d**. (**d**) Curvature of the flagellum at segments located at 15 µm (blue) and 25 µm (red) from the head. The curvature displays a sawtooth-like profile. (**e**) Power spectrum of the curvature at 15 µm and 25 µm. The fundamental frequency is $\omega_o = 20$ Hz.

The curvature profile shows the well-known bending wave propagating along the flagellum from the mid-piece to the tip (Fig. 1c). However, at a fixed arclength position, the curvature deviates in time from a perfect sinusoidal wave; instead, the curvature displays an asymmetric sawtooth-like profile in time, suggesting that multiple beat frequencies contribute to the overall waveform (Fig. 1d). Fourier analysis of the beat pattern reveals a fundamental beat frequency $\omega_o$ of approximately 20 Hz, but also higher-frequency components, mainly the second harmonic (Fig. 1e).

Principal-component analysis[18] provides further support for the presence of a second harmonic (Fig. 2). We decomposed the curvature profile $C(s,t)$ into normal modes $\Gamma_n(s)$ (Fig. 2a):

$$C(s,t) = \sum_n \chi_n(t)\Gamma_n(s), \quad (1)$$

where $\chi_n(t)$ is the amplitude of the n-th mode (Fig. 2b,c). The curvature is sufficiently well described using the first two modes (Fig. 2a), that account for about 90% of the signal (35 cells; Methods and Supplementary Fig. 2). The beat wavelength varied among sperm. However, after rescaling the arclength by the beat wavelength, the first two modes from different cells can be superimposed (Fig. 2a). Thus, the superposition of two eigenmodes recapitulates fairly well the beat pattern of human sperm.

The mode amplitudes $\chi_{1,2}(t)$ encode the time dependence of the flagellar shape. The probability $P(\chi_1,\chi_2)$ of observing a particular combination of mode amplitudes $\chi_1$ and $\chi_2$ adopts a typical limit-cycle shape (Fig. 2d)[18-20]. For a single-frequency cycle, the probability density would adopt an isotropic ring-like distribution, characterized by $(\chi_1,\chi_2) = a(\cos(\omega_o t), \sin(\omega_o t))$. However, the measured probability density displays two regions with higher probabilities at the "north" and "south" poles of the beat cycle, indicating the presence of a second harmonic (Fig. 2d). The average phase velocity for a given phase, $\langle \partial_t \alpha | \alpha \rangle$, reveals that the frequency smoothly oscillates between approximately 20 Hz and 40 Hz during each beat cycle (Fig. 2e). Thus, the beat pattern is indeed characterized by a fundamental frequency and its second harmonic.

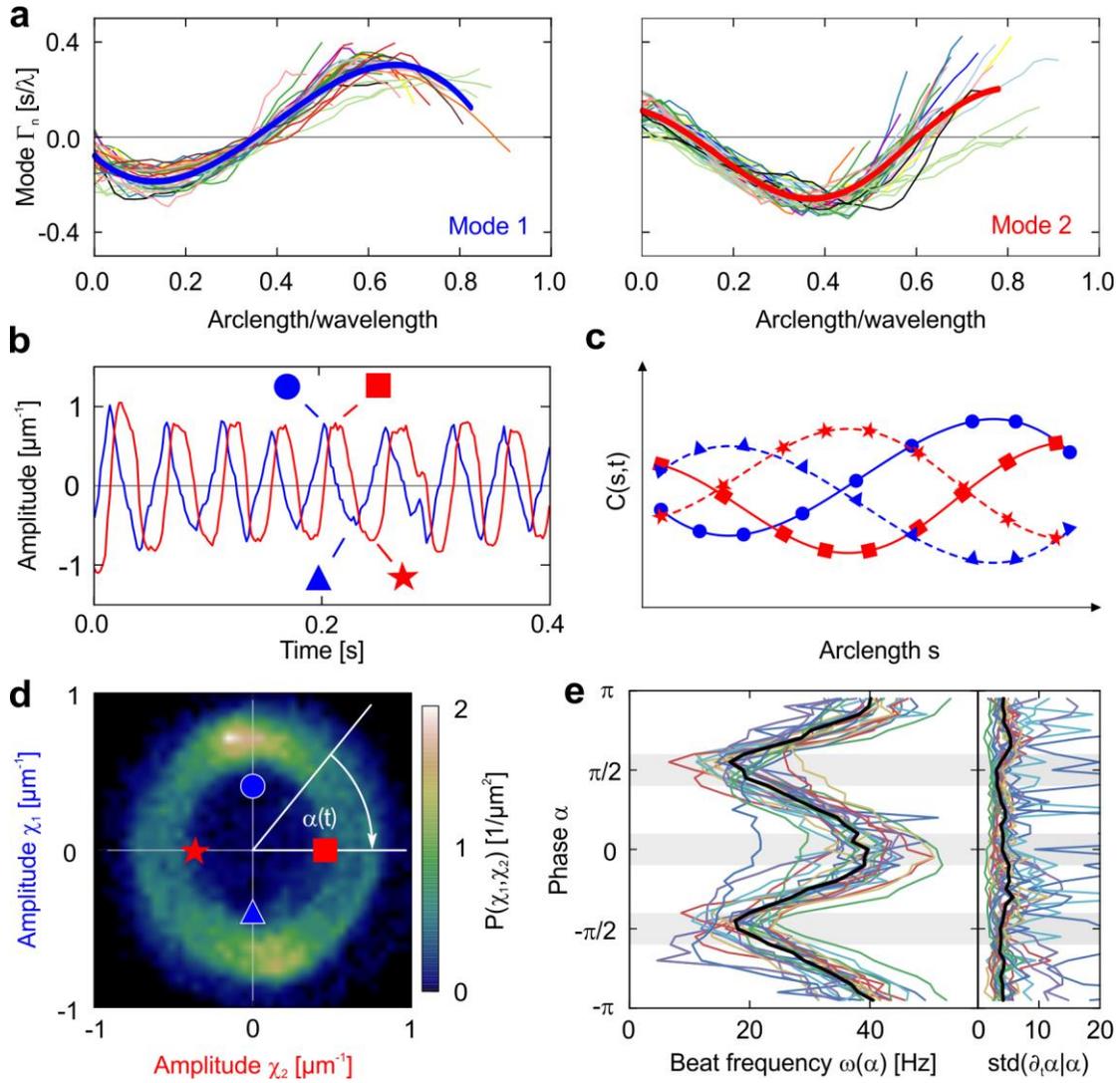

**Figure 2 | Principal-component analysis of the flagellar beat.** (**a**) Superposition of the two main normal modes of the flagellar beat for $n = 35$ human sperm cells. Each trace corresponds to a different cell. Although each cell has a different set of eigenmodes, when rescaled by the wavelength, the individual modes collapse onto a common curve (solid blue and red lines; see Methods). The wavelength appears longer than the flagellum, because it is traced only partially. (**b**) Time evolution of the mode amplitudes $\chi_1$ and $\chi_2$ for a representative recording. Of note, $\chi_2$ lags behind $\chi_1$ by a phase of $\pi/2$. The symbols (●,■,▲,★) indicate the modes shown in panel **c**, and correspond to the beat phase depicted in panel **d**. (**c**) Illustration of the composition of the beat by principal modes. The peak of the wave travels from left to right. (**d**) Histogram of the joint probability $P(\chi_1,\chi_2)$, averaged for 1.5 s from 26 different cells. During a full beat cycle, the phase $\alpha = \arctan(\chi_2/\chi_1)$ varies between 0 and $2\pi$. The symbols indicate the phase of the amplitudes in panel **b**. (**e**) Beat frequency at a fixed phase for each cell as $\omega(\alpha) = \langle \partial_t \alpha | \alpha \rangle$. The grey stripes highlight the phases of minimal and maximal frequency (20 Hz and 40 Hz, respectively). The black line indicates the median. The standard deviation of $\omega(\alpha)$ is nearly constant (right panel).

**A second-harmonic mode breaks the mirror symmetry temporally.** A planar beat with a single frequency becomes its mirror image after half a beat period $\tau/2$ (i.e. $C(s,t) = -C(s, t + \tau/2)$). Sperm using such a mirror-symmetric flagellar beat would swim on a straight path, and tethered sperm would not rotate. The second harmonic breaks this mirror symmetry; consequently, curved swimming paths and tethered-sperm rotation becomes possible. We examined theoretically how this broken symmetry generates the torque that drives rotation. With the "small-amplitude approximation"[2,21], the waveform of a flagellum oriented on average parallel to the x-axis can be described by a superposition of first and second harmonics:

$$y(x,t) = y_1 \sin(kx - \omega_o t) + y_2 \sin(kx - 2\omega_o t + \phi), \qquad (2)$$

where $k$ is the wave vector, $\phi$ is the phase shift between the two modes, and $\omega_o$ is the fundamental frequency. Note that equation (2) can be rewritten as $y(x,t) = Y(t)\sin(kx - \omega_o t + \Phi(t))$, where the amplitude $Y$ and phase $\Phi$ are functions of time. At any instant in time $t_o$, the shape $y(x, t_o)$ is still a sine wave, and is thus mirror symmetric in space. Therefore, no average flagellar curvature is produced. By contrast, because the amplitude $Y(t)$ is time-modulated, at any given point $x_o$ the temporal dependence of $y(x_o, t)$ is asymmetric in time. Such an asymmetric beat pattern allows steering, because the hydrodynamic drag forces during the two halves of the beat cycle do not cancel. For amplitude ratios $y_2/y_1 \lesssim 0.3$, $y(x_o, t)$ resembles an asymmetric sawtooth-like profile as in Fig. 2b.

The hydrodynamic drag on the flagellum is anisotropic, i.e. the drag coefficients in the perpendicular ($\xi_\perp$) and tangential directions ($\xi_\parallel$) are not equal. Each point along the flagellum is subjected to the drag force $\mathbf{f}(x,t) = -\xi_\perp \mathbf{v}_\perp - \xi_\parallel \mathbf{v}_\parallel$, where $\mathbf{v}(x,t) = (0, \partial_t y)$ is the velocity of the filament at time $t$ and position $x$ [2,7,22,23]. The net perpendicular force, averaged over one beat cycle is (see Supplementary Notes)

$$f_y(x) = \omega_o k^2 (\xi_\perp - \xi_\parallel) y_1^2 y_2 \cos(kx - \phi) \qquad (3)$$

In the presence of a second harmonic ($y_2 \neq 0$), the force $f_y$, integrated over the whole flagellum, does not vanish. The rotation velocity $\Omega$ around the tethering point is obtained by torque balance: the torque generated around the tethering point $T_a \approx \int_0^L x f_y(x) \, dx$ equals the viscous torque. For comparison with experiments, it is more useful to describe the waveform $y(x,t)$ in terms of local curvature $C(s,t)$, with amplitudes $C_1$ and $C_2$ instead of $y_1$ and $y_2$ in equation (2).

The general result for $\Omega$ (Supplementary Notes) depends on wavelength $\lambda$ and flagellum length $L$; for $\lambda \to L$, a simple relation is obtained:

$$\Omega = -\omega_o \frac{3L^3}{4(2\pi)^4} \frac{\xi_\perp - \xi_\parallel}{\xi_\perp} C_1^2 C_2 \sin\phi. \qquad (4)$$

Equation (4) illustrates that rotation results from the superposition of the first and the second harmonics coupled to the anisotropic drag (equation (2)). Note that $\Omega$ depends both on the amplitude $C_2$ of the second harmonic and on the phase shift $\phi$ between the two modes. We refer to $C_2 \sin(\phi)$ as the "second-harmonic intensity".

**Second-harmonic intensity and rotation velocity are correlated.** The experimental rotation velocity slowly varies with time (Fig. 1b), providing the means to test the predictions from equation (4). We determined the phase ($\phi$) and amplitude ($C_2$) of the second harmonic from the spectrogram of the flagellar curvature (Methods and Supplementary Fig. 3) and compared the second-harmonic intensity with the rotation velocity $\Omega$ (Fig. 3a). For each cell ($n = 35$), the correlation coefficient $R$ between the normalized rotation velocity $\Omega(t)/\omega_o$ and second harmonic intensity $C_2(t)\sin(\phi(t))$ was calculated by time averaging over the course of the experiment. To account for the approximations introduced for the derivation of equation (4), the phase $\phi(t)$ is corrected by a constant phase shift $\phi_o$ to yield $\phi_{\text{eff}}(t) = \phi(t) + \phi_o$. The constant shift is chosen such as to maximize the correlation coefficient $R$. We find that the second-harmonic intensity and the rotation velocity are highly correlated (Fig. 3a-c) ($R = 0.91 \pm 0.13$).

Alternatively, an average intrinsic curvature ($C_o$) of the flagellum might contribute to the rotation. An intrinsic curvature, which can generate an asymmetric beat, has been observed for some cilia and flagella[6,10,11]. The small-curvature calculation (Methods and Supplementary Notes) predicts that, for equal magnitudes of $C_o$ and $C_2$, both mechanisms contribute equally to the rotation frequency. However, the average intrinsic curvature of the flagellum is usually much smaller than the amplitude of the second harmonic ($|C_o|/|C_2| = 0.13$; Supplementary Notes and Supplementary Fig. 5). Therefore, we conclude that the second-harmonic contribution dominates. Accordingly, we find that the correlation of $C_o$ with the rotation frequency is weak ($R = 0.13 \pm 0.65$; $n = 35$). However, sometimes the average curvature, second harmonic intensity, and rotation velocity display a similar time course (Fig. 3a). In summary, these results support the hypothesis that human sperm steer with the second harmonic.

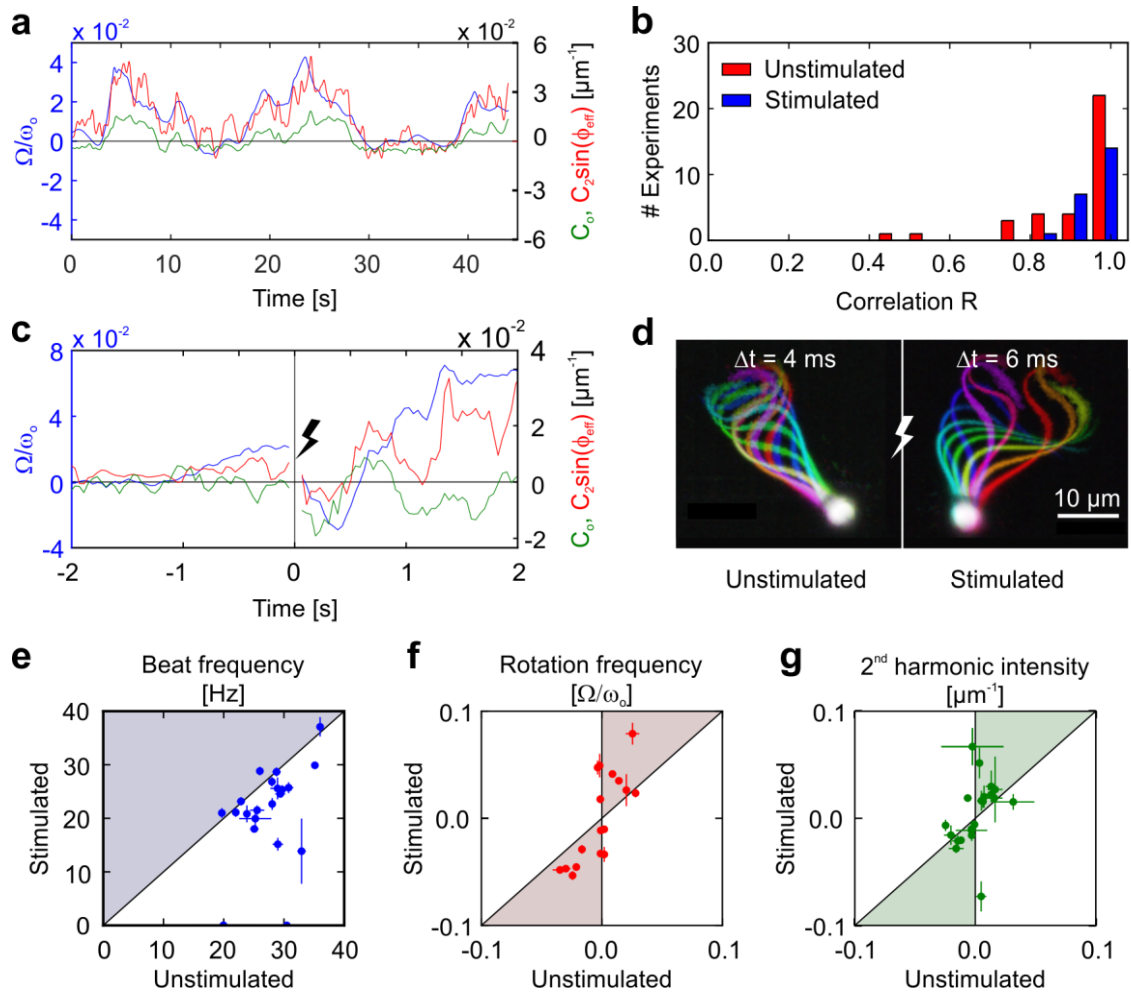

**Figure 3 | Second-harmonic intensity correlates with rotation velocity and is enhanced by progesterone.** (**a**) Normalized rotation velocity (blue line), second-harmonic intensity (red line), and average curvature (green line) for a representative sperm cell. (**b**) Histogram of the correlation $R(\Omega/\omega_o, C_2 \sin(\phi_{eff}))$. Red bars refer to unstimulated human sperm, blue bars to sperm stimulated with progesterone. We never observed anti-correlation ($R<0$). (**c**) Normalized rotation velocity (blue line) and second-harmonic intensity (red line), and average curvature (green line) 2.0 s before and after the release of progesterone with a flash of UV light (at $t = 0$). (**d**) Stroboscopic views of a sperm cell before (left) and after (right) stimulation with progesterone. Flagellar snapshots were recorded at $\Delta t = 4$ ms (left) and $\Delta t = 6$ ms (right) intervals. (**e-g**) Cell-by-cell comparison of the beat frequency (**e**), the rotation frequency (**f**), and the second-harmonic intensity (**g**), before and after progesterone release. Average values during 0.5 s before and after the stimulus. Points inside the colored areas correspond to an increase after the stimulation. Error bars are s.d.

**Changes in Ca$^{2+}$ concentration control second-harmonic intensity.** The flagellar beat of sperm and the steering response is controlled by changes in [Ca$^{2+}$]$_i$[12]. The female sex hormone progesterone evokes robust Ca$^{2+}$ entry into human sperm by activating the CatSper Ca$^{2+}$ channel[24,25]. The ensuing change in the flagellar beat pattern has been proposed to underlie hyperactivated motility and chemotaxis[26-29]. We used progesterone stimulation to examine whether $Ca^{2+}$ modulates the second-harmonic contribution. Sperm were imaged before and after photo-release of progesterone from a caged derivative (Fig. 3d)[14]. In Fig. 3e and f, we compare the beat pattern during 0.5 s before and after the release of progesterone. Although progesterone slowed down the beat frequency of human sperm (Fig. 3e), the rotation around the tethering point was enhanced (Fig. 3f). A direct comparison of the second-harmonic contribution before and after the release (Fig. 3c,g) demonstrates that progesterone modulates the second-harmonic intensity and, thereby, the rotation velocity; moreover, both measures are highly correlated (Fig. 3b,c,f,g). The strong second-harmonic component might thus represent the mechanism of hyperactivated beating of human sperm upon progesterone stimulation.

**An active elastic-filament model predicts that beating with two harmonics produces an intrinsic flagellar curvature.** Beyond a purely geometric description of the shape, we study by simulation the elasticity, forces, and the power generated or dissipated during a flagellar beat. A sperm cell is modeled as a tethered, actively beating filament of bending rigidity $\kappa$; hydrodynamic interactions are taken into account via anisotropic drag. The filament is driven by active bending torques $T(s, t)$, assuming a superposition of two traveling waves,

$$T(s,t) = T_1 \sin(ks - \omega_0 t) + T_2 \sin(ks - 2\omega_0 t + \psi). \qquad (5)$$

Due to hydrodynamic boundary effects, the phase shift $\psi$ of the torque can be different from the phase shift $\phi$ of the flagellar curvature in equation (4). All parameters in equation (5) and the bending rigidity $\kappa$ were derived by fitting to experimental data, including flagellar waveform, rotation velocity, and normal modes (Supplementary Notes). The simulation, which reproduces the beat pattern reasonably well (Fig. 4a, Supplementary Movie 1), provides several insights. First, constant torque amplitudes $T_1$ and $T_2$ along the flagellar arclength suffice to account for the experimental beat shapes, including the very high curvature of the end-piece. Thus, no structural inhomogeneity or differential motor activity along the flagellum is needed to account for this peculiarity of the flagellar beat shape. Second, although the bending forces are mirror-symmetric with respect to the filament displacement, a small average curvature is generated by the superposition of two harmonics that breaks the mirror symmetry of the beat waveform in both time (second harmonic) and in space (average curvature) (Fig. 4b,d). Third, the simulations confirm two predictions from equation (4): The rotation velocity $\Omega$ scales linearly with $T_2$ (Fig. 4c,d), and scales with the sine of the phase ψ (Fig. 4d, Supplementary Fig. 7). Fourth, for wavelength λ < L, the rotation velocity is largely independent of the wavelength; however, for longer wavelengths, the rotation velocity decreases (Fig. 4c). Finally, simulations of freely swimming sperm in 2D show that the curvature of the swimming path is controlled by the phase ψ (Fig. 4e), i.e. sperm could navigate by adjusting the phase ψ between the two harmonics.

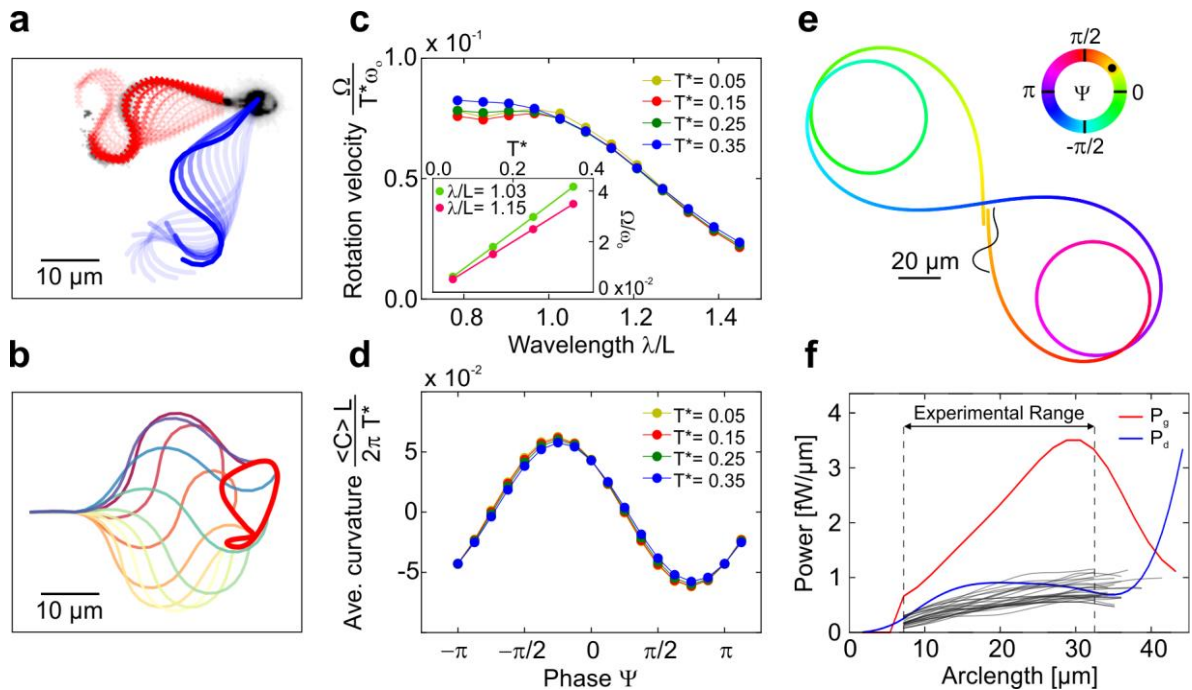

**Figure 4 | Simulations reproduce the beat and steering dynamics.** (**a**) Stroboscopic view of experimental (red) and simulated (blue) beat pattern using an active semi-flexible filament and anisotropic drag force. Time interval between snapshots (fading lines) is Δ$t$ = 2 ms. Simulation parameters: κ = 1.9 nN μm², $T_1$ ~ 0.65 nN μm, $T_2$ = 0.15$T_1$, ψ = 2.26, $ω_o$ = 30 Hz, $L$ = 41 μm, $ξ_⊥/ξ_∥$ = 1.81, $ξ_∥$ = 0.69 fNs/μm² and λ/$L$ = 0.65 (Supplementary Movie 1). (**b**) Representative simulation of flagellar beat with a second-harmonic amplitude $T_2$ = 0.3$T_1$. The mid-piece is aligned for visualization. The time interval between snapshots is Δ$t$ = 4 ms. The red thick line shows the non-symmetric trajectory of the flagellum tip. (**c**) Rotation velocity Ω versus normalized wavelength λ. Note that Ω has been normalized to the second-harmonic torque amplitude. The inset shows that Ω scales linearly with $T^*$ = $T_2/T_1$. (**d**) Average curvature <$C$> versus phase ψ of the second-harmonic torque. Note that the curvature has been normalized by $T^*$ = $T_2/T_1$. (**e**) Simulated sperm trajectory resulting from a slowly changing phase ψ over time (phase indicated by the color of the trajectory). By modulating the phase, sperm swim on curvilinear paths (Supplementary Movie 2). (**f**) Average dissipated power $P_d$ (blue) versus generated power $P_g$ (red) in simulations, and average dissipated power measured in experiments (grey lines). The simulated dissipated power shows good agreement with the experimental results. Of note, power is relocated along the flagellum.

**Energy consumption and dissipation.** Several aspects regarding the energetics of motile cilia and flagella have been studied, including traveling waves, power-and-recovery stroke, and metachronal waves[19,30-36]. For propulsion, not all beating gaits offer the same efficiency of energy consumption[37-39]. In fact, the flagellar beat pattern can be predicted from optimal swimming efficiency[38]. However, quantitative estimates of how power is used for bending and how power is dissipated along the flagellum of microswimmers are lacking[40-42].

Our simulations provide insight into the energetics of beating. Comparison of experimental results with simulations is only possible for power $P_d$ dissipated due to drag forces. The dissipated power $P_d$ (Supplementary Notes) increases along the flagellum within about 10 µm from the head, stays roughly constant for 25 µm along the entire principal piece, and then steeply rises towards the end piece, where the flagellum moves faster (Fig. 4f). The experimental recordings, which are restricted to a flagellar section between 7 and 35 µm from the head center, agree reasonably well with the simulations (Fig. 4f). From the simulations, we can also estimate the power $P_g$ generated by the instantaneous local torques described by equation (5) (Supplementary Notes). The generated power $P_g$ increases steadily, becomes maximal at about 30 µm down the flagellum, and decreases again towards the tip region. The distribution of generated and dissipated power differ along the flagellum: During the steady increase of $P_g$ to its maximum, the dissipated power $P_d$ stays almost constant; then $P_g$ quickly drops, whereas $P_d$ steeply increases thereafter (Fig. 4f). We conclude that the effects of local torques add up in order to generate large beating amplitudes and velocities in the tip region; by contrast amplitudes and velocities are smaller in the mid-piece region due to the tethering constraint and the head drag. Thus, although bending forces in eukaryotic flagella are locally generated along the length of the axoneme, power dissipation due to fluid drag is not equally distributed, yet relocated towards the tip region of the flagellum.

**Discussion**

Several mechanisms have been proposed that can produce asymmetric waveforms of the ciliary or flagellar beat. Symmetry breaking can emerge from structural features such as the central apparatus of the axoneme, or elastic filaments, or dynein motors that vary along the circumference or the long axis of the axoneme[43-45]. Here we identify a novel mechanism of symmetry breaking that is dynamic rather than static: two travelling waves of fundamental and second-harmonic frequency determine the beat asymmetry by their phase relation and the relative amplitude of each wave. In principle, this mechanism does not require a spatial or structural asymmetry that gives rise to an intrinsic curvature of the flagellum. Furthermore, simulations using homogeneous constant torque amplitudes along the flagellar arclength suffice to account for the experimental beat shapes (equation (5), Fig. 4a, Supplementary Movie 1). Hydrodynamic simulations suggest that intrinsic curvature of the midpiece affects swimming path curvature[10]. However, an intrinsic curvature and a dynamic component produced by the second-harmonic mechanism are not mutually exclusive. In fact, Fourier analysis of beat waveforms from different sperm species and *Chlamydomonas*[11] reveals a zero component or intrinsic curvature component[6] and at least two other components: a principal and a second harmonic[6,8]. However, in human sperm, the average curvature is small (Fig. 3a,c, Supplementary Fig. 5) and modulation of intrinsic curvature was not favored as a steering mechanism during rheotaxis[8].

Alternatively, buckling instabilities have been proposed to produce asymmetric beating[8,9]. These instabilities are enhanced at higher shear forces and flagellar compression[9]. However, progesterone stimulation slows down the beat frequency considerably, whereas the second-harmonic intensity and the rotation frequency are enhanced (Fig. 3e-g). These experiments, therefore, argue against dynamic buckling instabilities underlying mirror-symmetry breaking in human sperm.

Simulation of the flagellar beat shows that the second-harmonic intensity can control the swimming path of freely moving sperm and the rotation velocity of tethered sperm (Fig. 4e, Supplementary Movie 2). Furthermore, we find that progesterone-evoked $Ca^{2+}$ influx enhances the relative contribution of the second harmonic to the overall beat. Thus, the dynamics of principal and second-harmonic travelling waves could steer sperm across gradients of sensory cues that modulate the $Ca^{2+}$ concentration.

The mechanisms underlying the second-harmonic and its modulation by $Ca^{2+}$ are not known. However, dynein arms behave as endogenous oscillators that slide microtubules with a frequency set by the ATP concentration[46]. Thus, principal and second harmonics could be inherent properties of different dynein motors. Consistent with this idea, it has been shown that axonemal models from sea urchin sperm and flagella from *Chlamydomonas* mutants lacking the outer dynein arms beat at about half the frequency, indicating that inner and outer dynein arms could be tuned to produce different beat frequencies[47]. Furthermore, isolated *Chlamydomonas* flagella that were reactivated with varying ATP concentrations display beat amplitudes with two peak resonances at 30 and 60 Hz[48], and higher harmonics have been suggested to control steering during phototaxis of *Chlamydomonas*[49]. Future studies need to address the molecular mechanisms by which a second-harmonic mode is created and tuned for sperm steering.

**Methods**

**Sperm preparation**

Samples of human semen were from healthy donors with their consent. Sperm were purified by a "swim-up" procedure in human tubal fluid containing (in mM): 97.8 NaCl, 4.69 KCl, 0.2 $MgSO_4$, 0.37 $KH_2PO_4$, 2.04 $CaCl_2$, 0.33 Na-pyruvate, 21.4 lactic acid, 2.78 glucose, 21 HEPES, and 4 $NaHCO_3$; pH was adjusted between 7.3 and 7.4 with NaOH. After washing, human serum albumin (HSA, Scientific Irvine, USA; 3 mg/ml) was added, and sperm were incubated for at least 1 h at 37 °C and 10% $CO_2$ atmosphere.

**Sperm motility**

Single sperm cells were imaged in custom-made observation chambers of 150 µm depth. To gently tether the head of sperm cells to the glass surface, the HSA concentration in the buffer was reduced to 1 µg/ml, resulting in a large fraction of cells tethered to the surface with the head, but the flagellum was freely beating. The flagellar beat was recorded under an inverted microscope (IX71; Olympus) equipped with a dark-field condenser, a 20x objective (UPLANFL; NA 0.5), and additional 1.6x magnification lenses (32x final magnification). The temperature of the microscope was adjusted to 37 °C using an incubator (Life Imaging Services). Illumination was achieved using a red LED (M660L3-C1; Thorlabs), and a custom-made power supply. Images were collected at 500 frames per second using a high-speed CMOS camera (Dimax HD; PCO). For release of progesterone from its caged derivative (1 µM)[14] a brief flash (100 ms) of UV light was used (365-nm LED; M365L2-C; Thorlabs). UV light reached the sample through the backport of the microscope and a 380 nm long-pass dichroic filter (380 DCLP; Chroma). Tracking of the flagellum was achieved with custom-made programs written in MATLAB (Mathworks). The program identified the best threshold for binarization of the image by iteratively reducing the threshold until the expected cell area and coarse flagellar length in the image was achieved. This was followed by a skeleton operation to identify the flagellum. The position of the head was determined by fitting an ellipse around the tethering point.

**Rotation velocity**

The rotation velocity $\Omega(t)$ is obtained as the time derivative of the angle $\alpha$ between the x-axis and the vector connecting the head tethering point with the first tracked flagellar point. The angle $\alpha$ was filtered with a Gaussian of width 1 s to remove oscillations due to the fast beat.

**Flagellar curvature**

An arclength $s$ sampled every $\Delta s = 0.9$ μm from head to tip, is assigned to the tracked flagellum. Because the number of tracked flagellar points can differ from frame to frame (compare panels a and b in Supplementary Fig. 1), we analyse only the part of the flagellum that has been tracked for all frames. The curvature $C(s,t)$ is computed as the inverse of the radius of the circle that connects three contiguous points of the tracked flagellum, and is positive (negative) for counter-clockwise (clockwise) bends.

**Principal-component analysis**

The curvature $C(s,t)$ is a matrix of about 30 (in arclength) by 10,000 (in time) entries (Supplementary Fig. 1). We reduce the dimensionality of the dataset by principal-component analysis[18] to filter out white noise. The normal modes $\Gamma_n$ of the curvature are the eigenvectors of the non-standardized covariance matrix $M(s,s') = \langle C(s,t)C(s',t) \rangle_t$, with average curvature $\langle C(s,t) \rangle_t \approx 0$. Modes are sorted according to their eigenvalues $\sigma_n$. The number $n$ of eigenvalues $\sigma_n$ and eigenvectors $\Gamma_n(s)$ is equal to the number of tracking points ($n \approx 30$). We distinguish the relevant modes by comparing the eigenvalues of $M(s,s')$ with the eigenvalues of the correlation matrix of a random curvature $C_{ran} = \eta(s,t) \cdot \sigma$, where $\langle \eta(s,t)\eta(s',t) \rangle = \langle \delta(t-t')\delta(s-s') \rangle$ and $\sigma^2$ is the variance of the original curvature. Only the first three eigenmodes are statistically significant (Supplementary Fig. 2) and contribute with about 95% to the signal. The curvature $C(s,t) = \sum_{i=1}^{3} \Gamma_i(s)\chi_i(t)$ reconstructed from the first three modes and the corresponding amplitudes $\chi_i(t) = \int_0^L C(s,t)\Gamma_i(s)\,ds$ describes fairly well the experimental data (compare Supplementary Fig. 1a and 1b), and was used for further analysis.

To identify whether the same modes underlie all observed beat patterns (see Fig. 2), we rotate and mirror the first two eigenmodes, $\Gamma_1$ and $\Gamma_2$, of each experiment to maximize similarity with a reference pair of modes, $\Gamma_1^*$ and $\Gamma_2^*$. Because arclength is measured in wavelength units $\Delta s \rightarrow \Delta s / \lambda$, modes are interpolated to correct for the different tracking point density in the scaled representation. To avoid potential artefacts due to rotation and mirroring of modes, we use untransformed modes for other analyses.

**Curvature spectrogram**

We perform a discrete Fourier transform of the curvature $C(t, s = s_o)$ every 30 frames in the time window $(t_o - W/2, t_o + W/2)$ at fixed arclength $s_o$ and time window width $W = 250$ frames. $W$ is a compromise between time and frequency resolution. The peak at the fundamental frequency $\omega_o$ is clearly identified (Supplementary Fig. 3). The second-harmonic amplitude $C_2$ and phase $\phi$ are measured at $2\omega_o$.

**Correlation coefficient *R***

The correlation coefficient $R$ between the rotation velocity and second-harmonic intensity is defined as

$$R = \frac{\langle \Omega(t)/\omega_o \cdot C_2(t)\sin(\phi(t)+\phi_o)\rangle_t}{\sqrt{\langle (\Omega(t)/\omega_o)^2\rangle_t \langle (C_2^2(t)\sin^2(\phi(t)+\phi_o))^2\rangle_t}}, \quad (6)$$

where $\phi(t)$ is obtained directly by Fast-Fourier Transform of $C(t, s = s_o)$ (see Fig.1c). The phase offset $\phi_o$ is chosen such as to maximize $R$. Note that this is just one constant offset for each experiment with about 1,000 data points. We test whether the value of $\phi_o$ is independent of $s_o$ by comparing $R_{25}$, estimated from the curvature at 25 µm, with $R_{15}$, estimated from the curvature at 15 µm, using the phase estimated for $R_{25}$. The results are virtually identical (Supplementary Fig. 4). The correlation coefficient is centred around the values $C_2(t)\sin(\phi(t)+\phi_o) = 0$ and $\Omega(t)/\omega_o(t) = 0$. This choice agrees with the expectation that in the absence of a second harmonic the rotation velocity is zero.

**Author contributions:**
J.E, L.A, U.B.K, G.G designed research;
G.S, J.E performed data analysis and theoretical modeling;
J.F.J, L.A performed experiment and image analysis;
All authors discussed the results and wrote the paper.

**Acknowledgments:**

The authors declare no competing interests. Financial support by the Deutsche Forschungsgemeinschaft *via* the priority program SPP 1726 ''Microswimmers'' is gratefully acknowledged.

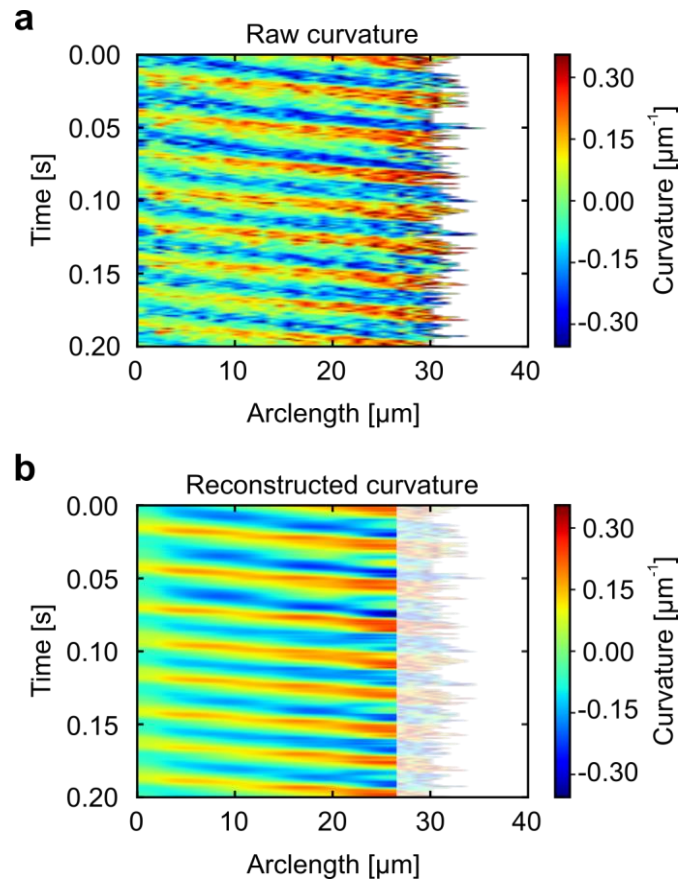

**Supplementary Figure 1.** (**a**) Unprocessed curvature of a beat flagellum derived from the tracking points. The number of points is not constant, hence some lines are longer than others. (**b**) The "reconstructed" curvature from the three most important normal modes of the beat pattern. The transparent part shows data that cannot be analysed with the proposed protocol. Arclength is measured from the first tracking point.

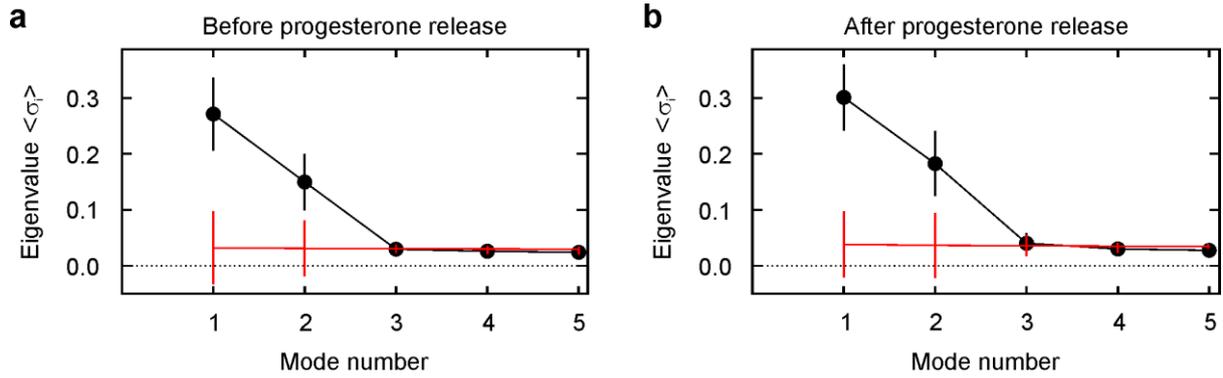

**Supplementary Figure 2.** Only the first three modes are statistically significant. (**a**) unstimulated sperm and (**b**) sperm stimulated with progesterone. Comparison of data (black) with random noise (red) shows that only the first three modes are significant. Values are mean ± s.d. ($n = 35$ (a) and $n = 26$ (b)). The third mode is barely above the threshold.

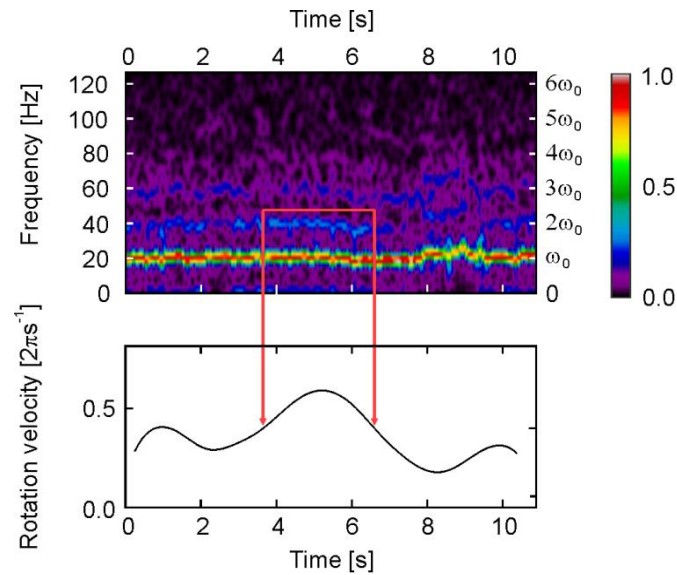

**Supplementary Figure 3. Upper:** Spectrogram of the curvature at arclength position $s_o \approx 25\ \mu m$ for one experiment. The power spectrum is color-coded and normalized to the maximum value. The fundamental mode corresponds to $\omega_o \approx 20$ Hz. Higher harmonics are observed at $2\omega_o$ and $3\omega_o$. Red lines indicate a window of particularly strong second-harmonic intensity. **Lower:** Rotation velocity. Note that the rotation velocity increases with the second-harmonic intensity.

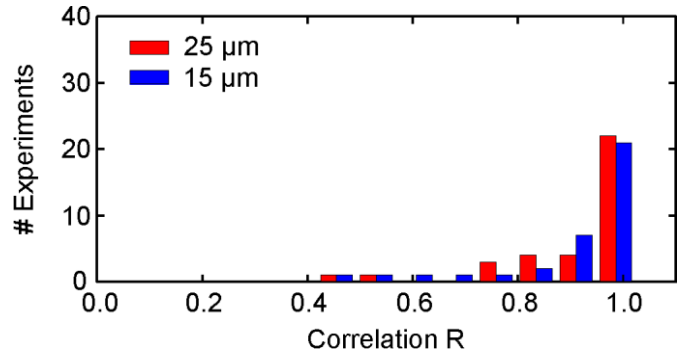

**Supplementary Figure 4.** Correlation coefficient *R* of rotation velocity and second-harmonic intensity for unstimulated sperm measured at arclength positions $s_o \approx 15$ µm and $s_o \approx 25$ µm.

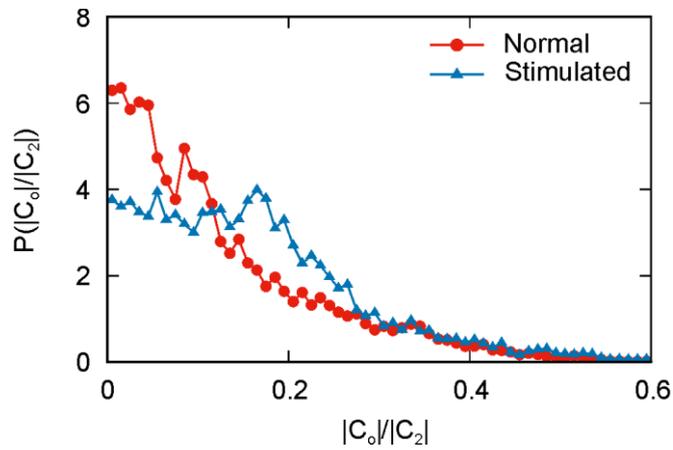

**Supplementary Figure 5.** Probability density for the ratio of average curvature $|C_o|$ and second-harmonic amplitude $|C_2|$ derived from experiments.

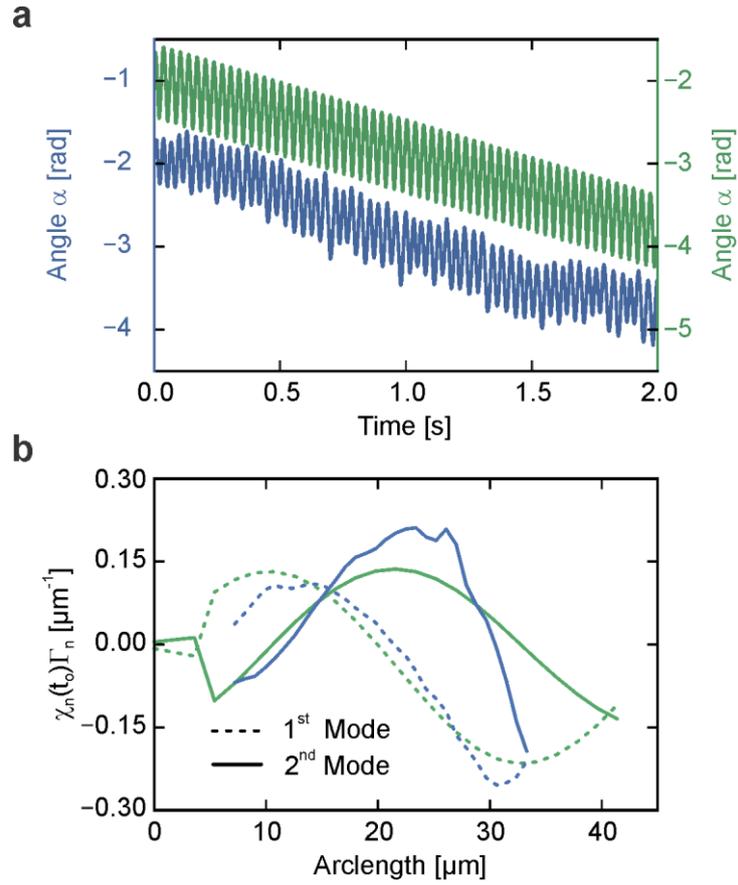

**Supplementary Figure 6.** Comparison between experimental (blue) and simulated (green) (**a**) rotation angle and (**b**) eigenmodes. For visibility, $t_o$ is chosen such that $\chi_n$ is maximal. Same simulation as in Fig. 4a in the Main Text and Movie 1. The simulation reproduces fairly well the angular velocity and the curvature eigenmodes.

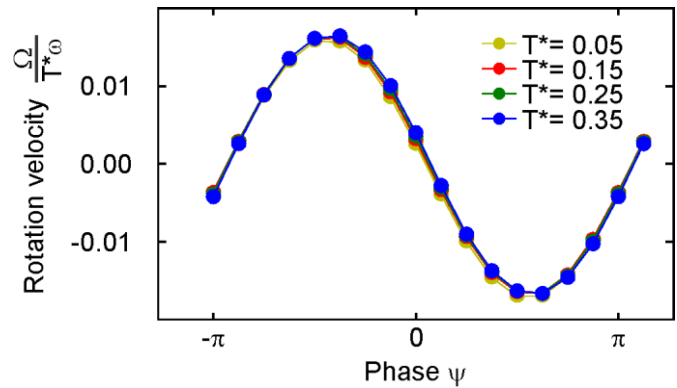

**Supplementary Figure 7.** The rotation velocity scales linearly with the applied torque ratio $T^* = T_2/T_1$ and trigonometrically with the torque phase $\psi$. Parameters are $\omega = 28$ Hz, $\kappa = 2.3$ nN μm$^2$, $\lambda/L = 0.6$, and $T_1 = 1.38$ nN μm.

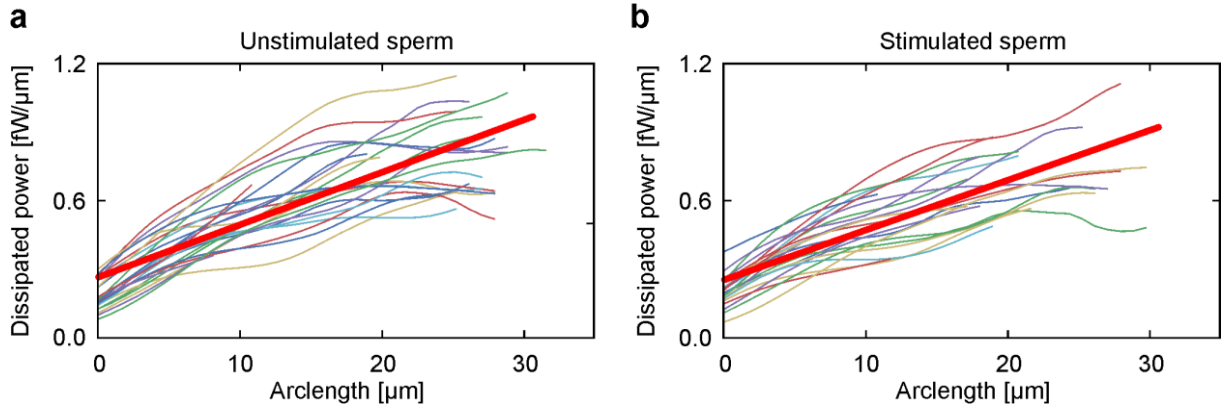

**Supplementary Figure 8**. Dissipated power density [fW/µm] for (**a**) unstimulated and (**b**) stimulated sperm. The drag ratio is $\xi_\perp / \xi_\parallel = 1.81$, with $\xi_\parallel = 0.69$ fNs$^{-1}$µm$^{-2}$. The red thick line is the interpolating average using equation (S20).

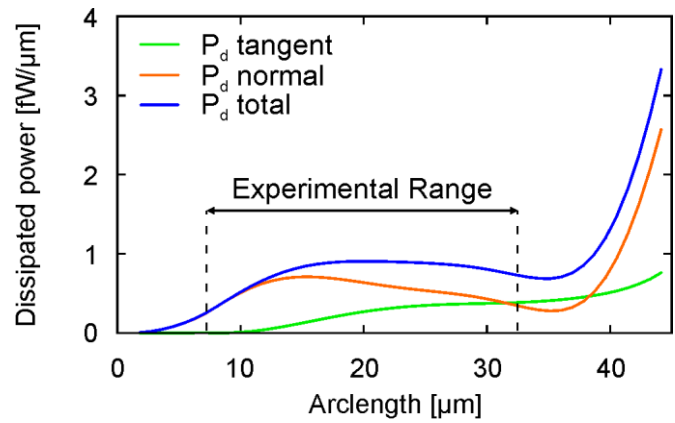

**Supplementary Figure 9.** Contributions to the total dissipated power $P_d$ by the tangential and normal components of the velocity.

**Supplementary Notes**

**Theory of rotation velocity due to the second harmonic**

We present here two approaches to derive the rotation velocity generated by the second harmonic. The first approach employs a small-amplitude approximation (equation (2) in the Main Text). This calculation demonstrates how the symmetry is broken and how it results in an average force in the direction perpendicular to the propagation direction. A second approach is based on a small-curvature approximation (equation (4) in the Main Text).

*Small-amplitude approximation*

For small deviations from a straight line, the flagellum is assumed to be oriented, on average, parallel to the x-axis. The deviation $y(x,t)$ from this line is given by the superposition of two harmonics,

$$y(x,t) = \epsilon \left[ y_1 \sin(kx - \omega_o t) + y_2 \sin(kx - 2\omega_o t + \phi) \right] \quad \text{(S1)}$$

with $0 < x < L$, $y_2/y_1 \leq 0.3$, and a small expansion parameter $\epsilon$.

The resistive-drag force is

$$\mathbf{f}(x,t) = -\xi_\parallel (\mathbf{v} \cdot \mathbf{t})\mathbf{t} - \xi_\perp (\mathbf{v} \cdot \mathbf{n})\mathbf{n}, \quad \text{(S2)}$$

where the velocity $\mathbf{v}(s,t)$, the tangent $\mathbf{t}(s,t)$, and normal $\mathbf{n}(s,t)$, vectors are given by

$$\mathbf{v}(x,t) = \begin{pmatrix} 0 \\ \partial_t y(x,t) \end{pmatrix}, \quad \mathbf{t}(s,t) = \frac{1}{N}\begin{pmatrix} 1 \\ \partial_x y(x,t) \end{pmatrix}, \quad \mathbf{n}(x,t) = \frac{1}{N}\begin{pmatrix} -\partial_x y(x,t) \\ 1 \end{pmatrix}, \quad \text{(S3)}$$

with normalization $1/N^2 = 1/(1+(\partial_x y)^2) \approx 1 - (\partial_x y)^2 + O(\epsilon^4)$.

Inserting equation (S3) into equation (S2), we obtain the instantaneous forces acting on the flagellum

$$\begin{aligned} f_x(x,t) &= (\xi_\perp - \xi_\parallel)\partial_t y \partial_x y + O(\epsilon^4), \\ f_y(x,t) &= -\xi_\perp \partial_t y + (\xi_\perp - \xi_\parallel)\partial_t y (\partial_x y)^2 + O(\epsilon^5). \end{aligned} \quad \text{(S4)}$$

The term $\xi_\perp \partial_t y$ in $f_y$ averages out during one period. The average net torque around the tethering point is then computed as

$$T_a = \frac{\omega_o}{2\pi} \int_0^{2\pi/\omega_o} dt \int_0^L dx\, x\, f_y(x,t) = -\epsilon^3 \omega_o y_1^2 y_2 (\xi_\perp - \xi_\parallel) \cdot \left[ kL\sin(kL - \phi) + \cos(kL - \phi) - \cos(\phi) \right]. \quad \text{(S5)}$$

In particular, for $\epsilon = 1$ and $2\pi/k \to L$

$$T_a \approx 2\pi \omega_o y_1^2 y_2 \sin(\phi)(\xi_\perp - \xi_\parallel) \quad \text{(S6)}$$

Supplementary Information

The torque $T_a$ generated by the second harmonic is balanced by the torque generated by the perpendicular viscous drag $T_v$. In line with the small-amplitude approximation, we estimate the viscous torque as the torque acting on a straight rod that is tethered at one end and rotates with angular velocity $\Omega$,

$$T_v = -\int_0^L dx\, \xi_\perp x(\Omega x) = \frac{\xi_\perp}{3}\Omega L^3. \tag{S7}$$

Torque balance ($T_v + T_a = 0$) finally yields

$$\frac{\Omega}{\omega_o} = \frac{\xi_\perp - \xi_\parallel}{\xi_\perp} \frac{6\pi}{L^3} y_1^2 y_2 \sin(\phi). \tag{S8}$$

*Small-curvature approximation*

Second, we consider a description of the flagellar shape in an expansion of small local curvature. This has the advantage that larger perpendicular deviation amplitudes can be included. The flagellar curvature $C(s,t)$ is written as:

$$C(s,t) = \delta\left[C_1 \cos(ks - \omega_o t) + C_2 \cos(ks - \omega_o t + \phi)\right], \tag{S9}$$

where $C_1$, and $C_2$ are the curvature amplitudes of the two harmonics and $\phi$ is the second-harmonic phase. Given the flagellar curvature $C(s,t)$ at time $t$, its spatial coordinates $\mathbf{r}(s,t)$ are given by

$$\mathbf{r}(s,t) = \mathbf{r}_o(t) + \int_0^s ds' \begin{pmatrix} \cos\psi(s',t) \\ \sin\psi(s',t) \end{pmatrix}, \tag{S10}$$

where $\psi(s,t) = \int_0^s ds'\, C(s',t) + \psi_o(t)$ is the local angle between the flagellar tangent and the x-axis. Here, we assume that sperm are clamped at their head such that $\psi_o(t) = 0$. By construction, the tangent vector $\mathbf{t}(t,s) = \partial_s \mathbf{r}(s,t)$ is normalized to unity.

The velocity of each line element along the flagellum is

$$\mathbf{v}(t,s) = \partial_t \mathbf{r}(s,t). \tag{S11}$$

For convenience, we rewrite the resistive force (equation (S2)) as

$$\mathbf{f}(s,t) = -\xi_\perp \left(1 + \zeta \hat{\mathbf{t}}\hat{\mathbf{t}}^T\right)\mathbf{v}, \tag{S12}$$

with $\zeta = (\xi_\perp - \xi_\parallel)/\xi_\perp$. If a net force is generated, it has to arise from the second term, which is proportional to the friction anisotropy $\zeta$.

The active torque around the tethering point is then obtained (to leading order in $\epsilon$) to be

$$T_a = \frac{\omega_o}{2\pi} \int_0^{2\pi/\omega_o} dt \int_0^L ds\, \mathbf{r}(s,t) \times \mathbf{f}(s,t) = -\frac{\omega_o \xi_\perp \zeta}{2\pi} \int_0^{2\pi/\omega_o} dt \int_0^L ds\, \mathbf{r}(s,t) \times \hat{\mathbf{t}}\hat{\mathbf{t}}^T \mathbf{v}(s,t)$$

$$= \delta^3 \xi_\perp \zeta \omega_o \frac{C_2 C_1^2}{8k^6}\left[\left(2k^2L^2 + 6\right)\sin(kL-\phi) - kL\cos(kL-\phi) + 6\sin(\phi)\right]. \tag{S13}$$

In the limit of $\lambda \to L$ and for $\delta = 1$, this simplifies to

$$T_a \approx -\xi_\perp \zeta \omega_o C_2 C_1^2 \pi \frac{4\pi\sin(\phi) + \cos(\phi)}{4k^6}. \tag{S14}$$

Note that the first term in the numerator is usually much larger than the second term. From torque balance and assuming a viscous torque as in equation (S7), we approximate $\Omega = 3T_a / \xi_\perp L^3$ by the expression given in equation (4) in the Main Text. Similarly, the average torque generated by an average flagellar curvature $C_o$ is obtained

$$T_a \approx \xi_\perp \zeta \omega_o C_o C_1^2 \pi \frac{\pi^2 - 3}{3k^6}. \tag{S15}$$

Equations (S14) and (S15) together with torque balance demonstrate that the rotation frequency is linear in both the second-harmonic amplitude $C_2$ and mean curvature $C_o$. Furthermore, for similar values of $|C_o|$ and $|C_2|$ the two terms contribute about equally.

**Comparison of second harmonic and average curvature**

Supplementary Fig. 5 shows the probability distribution of the ratio $|C_o|/|C_2|$. The second-harmonic contribution is always larger than that of the average curvature; the mean ratio is 0.13 for unstimulated and 0.16 for stimulated sperm. Because the ratio is much smaller than unity, the second harmonic dominates sperm rotation.

**Trajectory curvature**

The second harmonic generates a rotation of sperm around its tethering point. The rotation velocity $\Omega$ depends on the flagellar curvature amplitudes $C_1$ and $C_2$, the phase $\phi$, and the difference between drag coefficients, $\xi_\perp - \xi_\parallel$. We estimate how these parameters affect the trajectory of freely swimming sperm.

The time needed for sperm to complete a rotation around its center of rotation is $T = 2\pi/\Omega$ where $\Omega$ is approximately given by equation (S8) for tethered sperm. Assuming that freely swimming sperm have the same center of rotation, they move during the same time along a trajectory of length $vT$, with velocity $v = f_x / \xi_\perp$. Because $\Omega \ll \omega$, we can ignore the fast wiggling motion due to the beat.

The distance travelled in the time $T$ along a circular trajectory of curvature $C_{\text{traj}}$ is

$$2\pi / C_{\text{traj}} = vT = \frac{f_x}{\xi_\perp} \frac{2\pi}{\Omega}, \tag{S16}$$

where $f_x = \frac{\omega_o k}{2}(\xi_\perp - \xi_\parallel)(y_1^2 + 2y_2^2)$. Thus, for $\lambda \to L$ the trajectory curvature depends on the second-harmonic parameters as $C_{\text{traj}} \approx \frac{y_2}{1 + 2(y_2/y_1)^2} \sin\phi$. Therefore, the second-harmonic amplitude $y_2$ and phase $\phi$ determine the swimming trajectory (Fig.4e in the Main Text and Movie 2).

**Sperm model and simulations**

The sperm cell is modelled as a single semi-flexible filament, with one end fixed in its position by a stiff harmonic potential. The filament is discretized by beads separated by a distance $b$ between bead centres. Each bead is driven by three types of forces: bending, active, and viscous. The bending force is determined from the bending potential $U_b = \kappa/2b^3 \sum_{i=0}^{N-1}(R_{i+1} - R_i)^2$, where $R_i$ is the vector connecting the centres of bead $i$ and $i+1$, and $\kappa$ is the bending rigidity, which is obtained by fitting (see below). The active bending torque $T$ is given by equation (5) in the Main Text,

$$T(s,t) = T_1 \sin(ks - \omega_o t) + T_2 \sin(ks - 2\omega_o t - \psi). \tag{S17}$$

The viscous forces are modelled by an anisotropic drag. The tangential direction $\hat{t}_i$ at bead $i$ is defined as the line connecting beads $i-1$ to $i+1$, and the normal direction is computed by rotating $\hat{t}_i$ counter-clockwise by $\pi/2$. At the boundary beads, the tangential direction is identical to the bond direction. To reproduce the mechanics of the beat, the first 7.2 µm of the semi-flexible filament are taken to be inactive, thus mimicking head and midpiece (see Fig.4a in the Main Text). The equations of motion are integrated with an adaptive time-step Velocity-Verlet method.

The parameters were chosen as follows. A bond length $b = 1.8$ µm is a good compromise between number of points (hence computational performance) and accuracy. This implies that the flagellum is represented by a chain of approximately 25 beads (Supplementary Fig. 6b). The exact number of points depends on the flagellar length in the respective experiment.

Supplementary Information

The arclength is constrained locally by a harmonic bond potential of stiffness $k = 10$ nN / µm. We consider a system at low Reynolds number; therefore, the bead mass should not affect the dynamics. Nevertheless, it is useful to assign a small mass (11 pg) to each bead for a stable integration of the equations of motion. The ratio between the perpendicular and the parallel drag coefficients is chosen to be $\xi_\perp / \xi_\parallel = 1.81$, with $\xi_\parallel = 0.69$ fNs /µm$^2$, as measured for bull sperm[1]. The flagellar length, frequency, and wavelength are derived directly from the flagellar waveform. Subsequently, the bending stiffness $\kappa$ and the torque $T_1$ are adjusted to minimize the r.m.s. between the experimental and simulated principal modes (Supplementary Fig.6). A second fit adjusts $T_2$ and $\psi$ to the rotation velocity. The resulting stiffness $\kappa$ is typically about $\kappa \approx 2$nNµm$^2$ and thus compatible with known values for sea urchin sperm[2, 3]. Simulation results for beat patterns, rotation velocity, resulting average curvature, and power generation and dissipation are shown in Fig. 4 in the Main Text. In addition, Supplementary Fig. 7 shows that the trigonometric dependence of the rotation velocity $\Omega$ with respect to the torque phase $\psi$ agrees well with equation (S8). Note that Supplementary Fig. 7 compares the rotation velocity with respect to the driving torque ratio $T^* = T_2 / T_1$. Thus the simulations support the intuitive idea that $T_2$ has the same effect as $C_2$ on the rotation velocity.

**Generated and dissipated power**

The average power density dissipated by the viscous forces at arclength $s$ is

$$P_\mathrm{d}(s) = \frac{\omega}{2\pi} \int_0^{2\pi/\omega} dt\, \mathbf{f}(s,t) \cdot \mathbf{v}(s,t), \tag{S18}$$

where $f(s,t)$ is the resistive force equation (S2), and $v(s,t)$ is the filament velocity at position $s$ and time $t$. Because the torque is the force conjugate to the curvature, the average power generated by the torque is

$$P_\mathrm{g}(s) = \frac{\omega}{2\pi} \int_0^{2\pi/\omega} dt\, T(s,t) \partial_t C(s,t). \tag{S19}$$

The torque is given in equation (5) in the Main Text and in equation (S17) Thus, equations (S17), (S18), and (S19), can be used to estimate from experiments and simulations how the power is generated and dissipated along the flagellum.

*Experimental results*

Supplementary Fig. 8 displays $P_d(s)$ derived from the experimental data. The power dissipation increases approximately linearly towards the tip. For both unstimulated and stimulated sperm, a simple linear fit yields

$$P_d(s) = P_d^o + P_d^1 s, \qquad (S20)$$

where $P_d^o = 0.259 \pm 0.088$ [fW/µm] and $P_d^1 = 0.026 \pm 0.007$ [fW/µm$^2$].

*Simulation results*

In contrast to the experiments, the simulations allow a direct calculation and comparison of both the power dissipated in the medium and the power generated by the torques, without any further assumption and simplification. Figure 4f of the Main Text reveals a quite complex behavior. The generated power $P_g$ displays a maximum roughly 3/4 down the flagellar length, and decreases again towards the tip. In the regime of maximum power consumption, the dissipated power $P_d$ is nearly constant, but then increases sharply towards the tip. This implies that power is transported along the flagellum from the central part of the flagellum toward the tip. Interestingly, Supplementary Fig. 9 shows that in the front part and near the tip, dissipation is mainly due to motion of the flagellum perpendicular to the instantaneous contour, while motion tangential to the contour becomes increasingly important near the tip. These results can be understood qualitatively by considering that the effects of local torques add up to generate large beating amplitudes and velocities in the tip region, while amplitudes and velocities are small near the midpiece due to the tethering constraint.

*Discussion*

A direct comparison of experimental and simulation results is only possible for the dissipated power. Here, we find that the absolute values and the trends are very similar in the "experimental range" 10 – 30 µm of arclengths. This is not very surprising given that the beat shapes and beat frequencies from experiments and simulations are very similar (see Fig. 4a in the Main Text and Movie 1), and that the dissipated power is essentially a function of local beat velocities. The strong increase of dissipation near the tip observed in the simulations falls outside the experimental range.

A striking result of the simulations is the pronounced peak of the generated power in the central region of the flagellum. This behavior has important and interesting consequences. In particular, it implies that there is transport of power from the central portion of the flagellum

Supplementary Information

(20 – 35 µm) toward tip and midpiece. Thus, the power is not simply dissipated locally where it is produced. In the case of a swimming sperm, this should be even more pronounced, because part of the power generated by the flagellum is dissipated by the head. Hence, the mechanical structure of the flagellum mediates the power transport to other portions of the sperm cell. Further analysis is needed to gain deeper insights of power generation, thrust, and dissipation, as well as the differences between tethered and freely-swimming sperm.

**SI Movies**

*Movie 1. Beat pattern and sperm rotation*

Experimental recording, corresponding tracking (red), and simulation (blue) which best matches the first two eigenmodes of the curvature and the rotation velocity. The simulated and tracked flagella from previous frames are represented as fading lines ($\Delta t$ = 2 ms). After the first 210 ms, the movie is accelerated to show the rotation around the tethering point.

*Movie 2. Second-harmonic phase controls swimming direction*

The sperm trajectory in Movie 2 and the corresponding Fig. 4f of the Main Text are generated by slowly and linearly increasing the torque phase $\psi$ of the second harmonic from 0 to $2\pi$. The phase along the trajectory is indicated by colour. For a constant phase $\psi$, the trajectory is a simple circle.